\documentclass[%
 preprint,
 amsmath,amssymb,
 aps,
]{revtex4-2}

\usepackage{graphicx}
\usepackage{dcolumn}
\usepackage{bm}
\usepackage{color}
\usepackage{hyperref}

\begin{document}

\preprint{APS/123-QED}

\title{Unraveling friction forces of droplets on a non-wetting surface}

\author{Abhijit Kumar Kushwaha$^1$}
 \email{abhijitkumar.kushwaha@kaust.edu.sa}
 
\author{Sankara Arunachalam$^1$}%

 \author{Ville Jokinen$^2$}

\author{Dan Daniel$^1$}%

\author{Tadd T. Truscott}%
 \email{taddtruscott@gmail.com}
\affiliation{$^1$Mechanical Engineering, Physical Science and Engineering, King Abdullah University of Science and Technology\\
$^2$Department of Chemistry and Materials Science, School of Chemical Engineering, Aalto University, Espoo 02150, Finland}%

\date{\today}
\begin{abstract}
This paper explores the friction forces encountered by droplets on non-wetting surfaces, specifically focusing on superhydrophobic and superheated substrates. Employing a combination of experimental techniques, including inclined plane tests and cantilever force sensor measurements, we quantify friction forces across a broad range of velocities and surface types. Our results demonstrate that friction forces vary significantly with changes in droplet velocity and surface characteristics, transitioning from contact line pinning to viscous dissipation in the bulk of the droplet. We propose a universal scaling law that accounts for contact angle hysteresis, viscous dissipation, and aerodynamic drag, providing a comprehensive framework for understanding droplet dynamics on non-wetting surfaces. These findings offer valuable insights for optimizing surface designs in fluid transport and microfluidic applications, paving the way for enhanced efficiency and innovation in these technologies.

\end{abstract}

\maketitle
\newpage

\section{\label{sec:level1} Introduction}

Friction has ubiquitous influence across multiple disciplines- nature, engineering, and industrial processes. An ability to reduce friction has significant technological ramifications, affecting everything from large-scale fluid transport systems\cite{ceccio2010friction,aljallis2013experimental,park2014superhydrophobic} to precision microfluidic devices\cite{leslie2014bioinspired,sunny2016transparent,elzaabalawy2020potential,zhang2017robust}. This has led to a growing interest in the development of  non-wetting surfaces that can minimize friction. 
Non-wetting arises when a thin film of air is  trapped beneath the droplet and substrate which  leads to a substantial reduction in the effective contact area  between the substrate and droplet \cite{ou2004laminar,gogte2005effective} and consequently a reduction in friction \cite{vakarelski2011,lhuissier2013levitation,gauthier2016aerodynamic}. However, the forces that affect droplet friction change when considering the manner in which they interact with a surface. For instance, a droplet released on an inclined surface will experience aerodynamic drag, whereas a droplet pulled along a flat surface will not. Thus, unraveling the relevant forces affecting droplet mobility requires multiple experimental approaches and varied wettable surfaces. 


One type of non-wetting is formed by the Leidenfrost effect, where a thin vapor film  is generated beneath a droplet by gently depositing it onto a superheated surface. The insulating cushion of vapor layers formed beneath the droplet causes it to hover over the substrate without wetting \cite{leidenfrost1966fixation}. This phenomenon arises when the substrate temperature exceeds the critical temperature ($T>300^\circ C$ when the droplet is pure water), commonly identified as Leidenfrost temperature, for which the evaporation time of the droplet is maximized \cite{leidenfrost1756}. The hovering droplet exhibits extreme mobility as the insulating vapor layers remove solid-liquid contact, giving rise to zero adhesion and vanishing droplet friction \cite{linke2006,lagubeau2011,vakarelski2011,bourrianne2019cold}. 
Recent studies \cite{lagubeau2011,bouillant2018} have demonstrated that a Leidenforst droplet can exhibit characteristics of rolling and slipping, and even self-propulsion if the flow morphology is asymmetric. 

Extreme mobility and reduced friction is not only limited to a superheated surface. It can also arise at ambient temperature for a liquid droplet rolling or sliding on a substrate exhibiting superhydrophobicity \cite{mouterde2019superhydrophobic,backholm2020water}. A superhydrophobic surface comprises nano/micro-scale protrusions which entraps air to maintain a stable air layer, reducing the effective solid/liquid contact area. These surfaces are characterized by high-contact angles with low hysterisis which reduces  contact-line friction \cite{rothstein2010}. When a droplet rolls or slides over the superhydrophobic surface with increasing velocity, the friction emanates not only from the contact-line pinning-depinning process but also from the viscous dissipation that arises by the circulatory flow in the vicinity of the contact disc between the droplet and the substrate \cite{olin2013water,mouterde2019superhydrophobic,backholm2024toward}. Herein, we combine these two friction forces and refer to them as viscous dissipation. A superhydrophobic surface can also give rise to an aerodynamic Leidenfrost effect by entrapping air when moving sufficiently \cite{gauthier2016aerodynamic,sawaguchi2019droplet} which means that the viscous forces can potentially be reduced.

Aerodynamic drag (not to be confused with aerodynamic Leidenfrost) becomes significant as the velocity of the droplet increases\cite{mouterde2019superhydrophobic}.  
They argue that aerodynamic drag begins to dominate resistive forces as the droplet approaches terminal velocity. Either way, the dynamics of a droplet racing down a gently inclined substrate (Leidenfrost or superhydrophobic surface) is governed by three main factors:  viscous dissipation from droplet curvature near the substrate \cite{daniel2017oleoplaning,keiser2020universality}, hydrodynamic viscous dissipation in the bulk \cite{mahadevan1999rolling}, and the aerodynamic resistance of the droplet \cite{mouterde2019superhydrophobic}. These forces are balanced by the external force of gravity. 

A cantilever force sensor has recently been used to measure the friction forces of a droplet moving on flat superhydrophobic surfaces 
\cite{daniel2017oleoplaning,daniel2018origins,backholm2020water}. A scaling law for friction has been proposed which shows that the friction scales as $(Ca=\eta U/\sigma)^{2/3}$\cite{daniel2017oleoplaning,keiser2017drop,keiser2020universality}, where $Ca$ is the capillary number, $\eta$ is the viscosity of the film, $U$ is the velocity, and $\sigma$ is the surface tension of the droplet. However, these studies focused on a droplet moving over a lubricant-impregnated surface, specifically a thin oil film with much higher viscosity than air. Whereas, \citet{backholm2024toward} looked at a water droplet moving above a critical velocity on a superhydrophobic surface with a thin air-film trapped underneath the droplet. 
Due to the low velocities associated with these studies the aerodynamic drag and viscous dissipation are typically ignored, which must be included at higher velocities. Thus, a comprehensive scaling law that explains friction over a wide range of velocities and capillary numbers remains elusive. To address this gap, we conducted a systematic study, examining friction force as a function of velocity over six orders of magnitude (from $10 \mu$ms$^{-1}$ to 2 ms$^{-1}$). We are able to show transitions between dominant friction forces as the capillary number is increased. For low capillary numbers friction is constant and depends on contact angle hysteresis. For intermediate capillary numbers, friction force scales as $Ca^{2/3}$, similar to lubricant-impregnated surfaces. For large capillary numbers we find that viscous dissipation begins to play a larger role and the aerodynamic drag is small enough to ignore (for the velocities presented). We propose  a scaling law that is based on all of the relevant forces rather than just one. Additionally, we investigate how friction correlates with variations in air-film thickness beneath the droplet. 
 
 

\section{\label{sec:level2}Experimental setup}
\begin{figure*}
\includegraphics{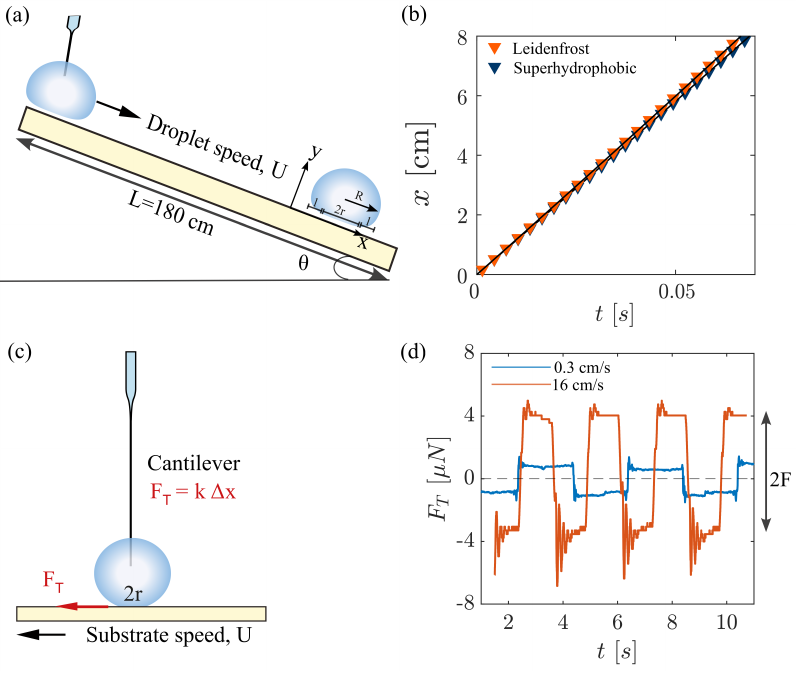}
\caption{\label{fig:1}Schematics of the experimental setup showing two different approaches to estimate the friction force. (a) A force balance approach where a droplet moves on an inclined ($\theta$) stationary non-wetting surface. (b) Friction measurement by a cantilever force sensor where the droplet attached to the cantilever rests on the superhydrophobic substrate while the substrate moves back and forth. (c) $x$-centroid position of the droplet along the plate from camera data (10 cm field of view). (d) The total friction force as a function of time, measured by the cantilever force sensor for two different substrate speeds. }
\end{figure*}
  
We use two different experimental methodologies to measure friction for a droplet moving on non-wetting surfaces: a force balance approach in which a droplet slides down an inclined non-wetting stationary surface (see Fig. \ref{fig:1}~a \& b), and a cantilever force sensor approach in which the droplet is stationary while the flat and rigid non-wetting surface moves (c \& d).

\subsection{\label{sec:exp1}Force balance approach}

The inclined plane experimental setup (Fig. \ref{fig:1}a) consists of three main components: (i) a droplet dispensing system, (ii) a long non-wetting inclined surface on which the droplet slides, and (iii) a high-speed imaging system. A blunt dispensing needle connected to a syringe pump via surgical tubing dispenses a Milli-Q water droplet of 42 $\mu$l onto one end of a non-wetting surface. The non-wetting surface sits on a custom-built platform mounted to a motorized vertical translation stage (ThorLabs MLJ250) to adjust the tilt angle ($\theta$), consequentially the velocity.  The non-wettability characteristics of the surface is achieved by make it either superhydrophobic or superheated. For superhydrophobicity, we spray a 1.8 meter long transparent glass plate with Glaco Mirror Coat Zero, a commercially available colloidal suspension of hydrophobic silica nano-beads in isopropanol. Black silicon surface was made by cryogenic deep reactive ion etching as described in \citet{Koh2024}. 
We also make a copper surface non-wettable by heating it above the Leidenfrost temperature ($T>300 ^\circ C$). The time-resolved images of a droplet sliding down both surfaces (Fig.\ref{fig:2}) is acquired using a high-speed camera (Phantom T3610) fitted with 180 mm lens at 5000 fps with a field of view of 10 cm. A 200 watt LED light (Godox 200 SL II) is used as a backlight. Fig \ref{fig:1}(c) shows the $x$-centroid of a droplet sliding on a superhydrophobic and a Leidenfrost surface as a function of time. We compute the velocity and acceleration by fitting a second-order polynomial to the $x$-centroid data.    

\subsection{\label{sec:exp2}cantilever force sensor approach}
Fig. \ref{fig:1}(b) shows the cantilever force sensor setup. A 7 cm long cantilever made of an acrylic capillary tube with an inner (0.288 mm) and outer radii (0.360 mm) is positioned vertically above the surface. The tip of the cantilever is submerged in a droplet (20 or 30 $\mu$l) resting on the substrate. The substrate is mounted on a horizontal-motorized translation stage (Thorlabs DDS220), allowing it to move at velocities ranging from  10 to 3$\times10^5 \mu$ms$^{-1}$. The deflection of the cantilever ($\Delta x$) is captured at frame rates from 100 to 1000 fps using a Kronos camera fitted with a $6.5\times$ zoom lens. We use open-source software (Tracker) to quantify the deflection. Having determined $\Delta x$, we directly estimate the total friction as $F_T = k \Delta x$, where $k$ is the stiffness of the cantilever; determined similar to the approach of \cite{backholm2019micropipette, daniel2017oleoplaning}. 
Figure \ref{fig:1}(d) shows the friction values as a function of time computed directly from the calibrated cantilever sensors for two different substrate velocities.

\subsection{\label{sec:RICM} Interferometry measurement}
We performed  Reflective Interference Contrast Microscopy (RICM) experiments to visualize the changes in the air-film thickness between the droplet and superhydrophobic surface. The setup includes a LED light emitting 530 nm green light focused onto the surface through a microscope objective (2$\times$ magnification and numerical aperture NA = 0.055). Interferometric recordings of the thin air-film interference patterns are captured at 3000 fps using a Kronos camera attached to an infinity corrected tube lens. Separately, we obtain absolute thickness of the air-film (point measurement) with white-light interferometry in which a co-axial illumination fibre optic cable (RP23 Thorlabs) shines white light onto surface from below. The fibre optic cable also consists of multi-mode read fibre which directs the reflected light into specterometer (CCS100 Thorlabs). The absolute local film-thickness is then deduced by analyzing the peaks and valleys of the normalized spectroscopic signal acquired from the spectrometer.

\begin{figure*}
\includegraphics{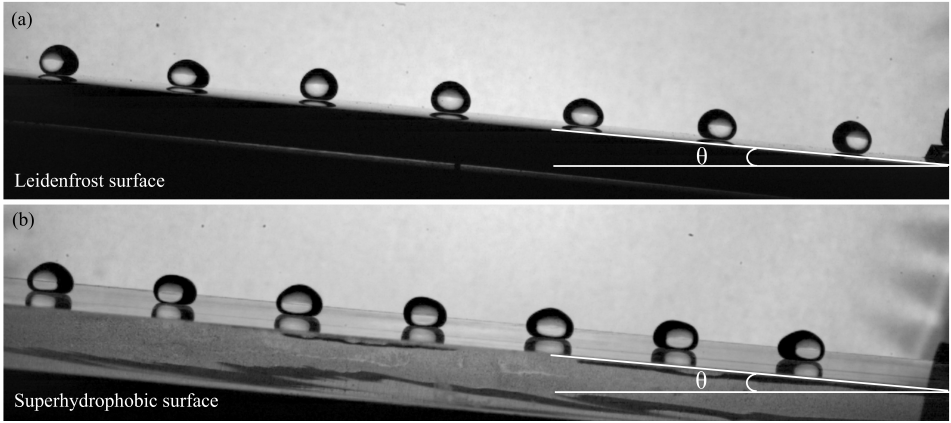}
\caption{\label{fig:2} Seven super imposed images of a 42 $\mu$l water droplet moving on an inclined superheated surface with a total length $L = 125~ $cm (a) and superhydrophobic surface with a total length $L = 180 ~$cm (b). Both non-wetting surfaces are inclined at $\theta=5$\textdegree. The field of view of the camera shows only the bottom most 11 cm of the surfaces. The superimposed droplet images are separated by 1.5 ms. The droplet on the Leidenfrost surface is moving at $U=1.2$ ms$^{-1}$ while the droplet on the superhydrophobic surface is moving at $U=1.17$ ms$^{-1}$. }
\end{figure*}

\section{\label{sec:level2}Results and discussions}
We begin by showing the sequential motion of a 42 $\mu$l water droplet moving on two different non-wetting surfaces: superheated surface and superhydrophobic surface. This sequential motion is visualized by overlaying experimentally captured side-view images of the droplet as it moves down the $\theta=5$\textdegree inclined surface (Fig. \ref{fig:2}). Seven superimposed images of a water droplet moving down a 125 cm long superheated surface is shown in Fig. \ref{fig:2}a and a 180 cm long superhydrophobic surface in Fig. \ref{fig:2}b (camera field of view is 11 cm from the end of the plate for both cases),  both tilted at $\theta=5$\textdegree. Images are spaced by 1.5 ms. A qualitative comparison of the droplet motion reveals that the droplet moving on the superheated surface attains a higher velocity compared to that of the superhydrophobic surface, despite traversing a smaller distance on the superheated surface. We quantitatively examine the resulting kinematics of a droplet as it moves down an inclined superheated or superhydrophobic surface. The total friction force is  then explored by comparing both inclined plate and cantilever measurements.

\subsection{\label{sec:level22}Droplet kinematics}

\begin{figure*}
\includegraphics{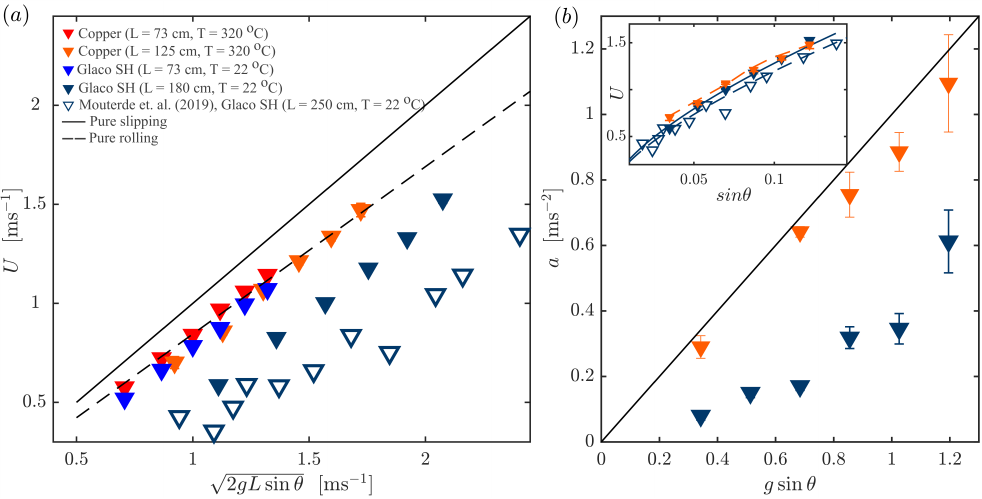}
\caption{\label{fig:3}(a) Measured velocity comparisons of a water droplet ($42 \mu l$) moving on inclined superheated and superhydrophobic surfaces versus the maximum potential energy velocity. A comparison is made against the theoretical pure slipping ($U_s = \sqrt{2gL\sin\theta}$) and pure rolling ($U_r = \sqrt{(10/7)gL\sin\theta}$) velocities of a solid spherical body. Data extracted from \cite{mouterde2019superhydrophobic} as marked. (b) Measured acceleration versus gravitational acceleration (line) for two cases in (a). Inset shows $U$ is a function of $sin\theta$ where the line comes from Eq.\ref{eq:2}.}
\end{figure*}

  
In Fig. \ref{fig:3}(a), we compare the experimentally observed velocity of a droplet moving on a tilted superheated surface and a superhydrophobic surface alongside two cases of a theoretical solid sphere perfectly rolling and perfectly slipping. The comparison is made for three different lengths ($L=73$, 125, \& 180~cm) of inclined plates and six different tilt angles ranging from $\theta$ =  2 to 6\textdegree. The pure slipping and rolling velocity of a solid sphere moving over an inclined plane of length L is defined, respectively, as $U_s =\sqrt{2gL\sin{\theta}}$ and $U_r = \sqrt{(10/7)gL\sin\theta}$. We notice that the velocity of the Leidenfrost droplet is closer to that of a solid body engaged in pure rolling motion than pure slipping. One might expect the velocity to be more like slipping, however, others have observed that a Leidenfrost droplet has rotational motion when moving or stationary \cite{bouillant2018}. Fig. \ref{fig:3}(b) shows the acceleration of each case and the inset shows variation in the velocity as a function of $sin\theta$
. We also compare the velocity of a water droplet on an inclined superhydrophobic plate from \citet{mouterde2019superhydrophobic} in Fig.\ref{fig:3}(a). Although the plate is longer (250 cm) and the droplet is larger (100 $\mu$l) the terminal velocity of the droplet is smaller than our cases. This is likely due to the different surface characteristics. 



A droplet racing down a superhydrophobic inclined surface (L = $73$~cm) shows a velocity profile that closely resembles that of the Leidenfrost droplet for all $\theta$, as indicated in Fig.~\ref{fig:3}(a). However, further down the plate ($L=180$ cm) the droplet velocity is lower for all $\theta$. This reduction in velocity suggests that the droplet has reached near-terminal velocity, indicating that its kinematics are influenced not just by gravity but also by friction resulting from viscous dissipation in the droplet and the thin air-film trapped between the droplet and the substrate, along with aerodynamic resistance. Furthermore, the acceleration is much lower than than the acceleration of gravity (Fig. \ref{fig:3}b), as well as lower than the Leidenfrost surface. This mainly indicates that the droplet on the superhydrophobic surface is at or near terminal velocity. 

\subsection{\label{sec:level22}Total friction force}

\begin{figure*}
\includegraphics{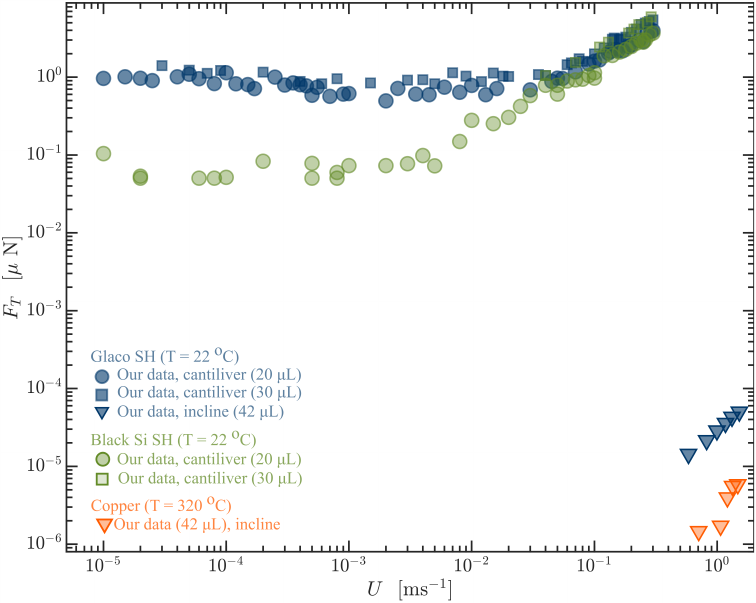}
\caption{\label{fig:4}Total fiction force plotted versus velocity, measured on various non-wetting surfaces: Glaco-coated superhydrophobic surface, black-silicon surface, and a Leidenfrost surface. The velocity corresponds to the speed of a droplet moving down an inclined plane when friction force is estimated using forced balance approach. For measurements with a cantilever force sensor, velocity refers to the speed of the substrate.}
\end{figure*}

The total frictional force on a droplet depends on both the shape and size of the droplet. A droplet of volume $V$ with density $\rho_w$, viscosity $\eta_w$, and surface tension $\sigma$ can deform due to gravity when placed gently on a non-wetting surface. The degree of deformation depends on the whether the nominal radius ($R_o = (3V/4\pi)^{1/3}$) is smaller or greater than the capillary length ($l_c=(\sigma/\rho g)^{1/2}$) and is governed by the Bond number ($Bo = (R_o/l_c)^2$)\cite{mahadevan1999rolling}. A sufficiently small droplet with $Bo\ll1$ retains a quasi-spherical shape. Whereas, a droplet becomes more flattened when the droplet radius approaches the capillary length ($Bo\approx1$). The droplet deformation at the bottom forms an assumed radius ($r$) near the substrate (shown in Fig.~\ref{fig:1}b). The viscous dissipation for such a droplet occurs at the length scale of $\delta=(\eta r/\rho U)^{1/2}$ which is the viscous boundary layer within the droplet \cite{mahadevan1999rolling}. Note, the maximum radius of the flattened droplet is given by $R \approx (V/\pi l_c)^{1/2}$.

The friction forces in the cantilever versus inclined plane experiments are not the same. In the case of our cantilever measurements, a superhydrophobic surface moves underneath a droplet, where its motion is opposed by friction forces arising from: hydrodynamic viscous dissipation in the bulk which arises from interaction of the air film \cite{mahadevan1999rolling}, viscous dissipation from the droplet curvature near the substrate \cite{daniel2017oleoplaning,keiser2020universality}, and the pinning/de-pinning process when no air-layer is present (contact angle hysteresis). The forces on the droplet in the cantilever case then become 
\begin{equation}
    F_{T}= 
\begin{cases}
    2r\sigma(\cos\theta_r - \cos\theta_a),& \text{if not levitated}\\
     \alpha \frac{\eta_w U}{\delta}\pi r^2 + \beta_1\Big(\frac{\eta_a U}{\sigma}\Big)^{2/3}2\pi\sigma r,              & \text{otherwise}
\end{cases}
\label{eq:1}
\end{equation}
We find $\alpha = 0.12$ and $\beta_1 = 6\pi$ as a best fit for all cantilever cases including \cite{daniel2019hydration}. Whereas, in the case of an inclined superhydrophobic and superheated surface, external gravitational forces drive the motion and aerodynamic drag \cite{mouterde2019superhydrophobic} forces are also considered such that
\begin{equation}
F_T = mg\sin\theta - ma=\alpha\frac{\eta_w U}{\delta}\pi r^2 + \beta_2\Big(\frac{\eta_a U}{\sigma}\Big)^{2/3}2\pi\sigma r + \zeta\frac{\rho_aU^2}{Re^{1/2}}\pi R^2.
\label{eq:2}
\end{equation}
where $\alpha=0.12$, $\beta_2=1$, and $\zeta = 1$ for all cases presented here including \cite{mouterde2019superhydrophobic}. These pre-factors are determined by fitting $U$ as function of $sin\theta$ in the inset of Fig.\ref{fig:3}. 

Fig. \ref{fig:4} illustrates the variation in total friction as a function of velocity for two different types of superhydrophobic surfaces in the cantilever experiment: (i) a Glaco-coated surface, and (ii) a black silicon surface. For these cases the total friction does not significantly increase despite a three-order magnitude rise in velocity from $U = 10^{-5}$ to $10^{-3}$~ms$^{-1}$. This consistent friction force across a broad range of velocities suggests that the friction primarily arises from contact-angle hysteresis similar to \cite{backholm2020water,backholm2024toward} and is velocity-independent (Eq.\ref{eq:1}). With a further increase in the velocity $U>10^{-2}~$ms$^{-1}$, we find that the friction forces steadily increase for the black-silicon surface at lower velocities than for the Glaco-coated superhydrophobic surface. The transition for both surfaces occurs when an air-layer forms below the droplet sufficient enough to eliminate the contact line pining for most of the droplet as shown in \S\ref{sec:interferometry}. 
The functional dependence of friction on velocity at these increased speeds, implies that the contact-line friction is no longer the dominant friction force driving the dynamics, and the black-silicon surface allows the droplet to levitate more easily than the Glaco-coated superhydrophobic surface. 
In the cantilever experiments, the substrate is moved while the droplet remains stationary, ruling out any aerodynamic resistance. Thus, the additional friction likely comes from viscous dissipation in the bulk and along the rim. 
Finally, we also observe a small increase in the friction force associated with an increase in droplet volume. 


In the case of a droplet on an inclined plate, a monotonic increment in the friction with velocity is also observed in Fig. \ref{fig:4}.  The friction stemming from the contact line pinning does not contribute to the total friction as the droplet has levitated (aka aerodynamic Leidenfrost) which we will show in detail in section \S\ref{sec:interferometry}. Further, the friction force experienced by a droplet on an inclined plane are five orders of magnitude smaller than in the cantilever experiments while the droplet on a superheated surface (Leidenfrost) is another order of magnitude smaller than the superhydrophobic surface.  



\begin{figure}
\includegraphics{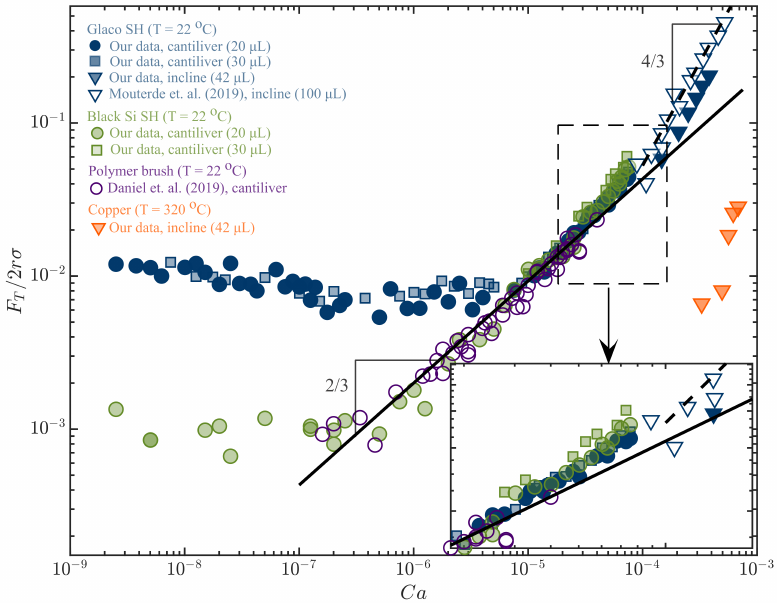}
\caption{\label{fig:5} (a) The non-dimensional friction force as a function of capillary number for cantilever and inclined plate experiments. Low $Ca\lesssim10^{-6}$ have constant friction forces associated with contact line friction. Moderate $10^{-6}\lesssim Ca \lesssim 5\times10^{-5}$ the friction from the droplet curvature near the substrate scales with $Ca^{2/3}$. Whereas larger $Ca\gtrsim10^{-4}$ deviate with droplet size or surface type.  Other studies as marked \cite{mouterde2019superhydrophobic, daniel2019hydration}. 
}
\end{figure}

The variation of non-dimensional friction as a function of capillary number in Fig. \ref{fig:5} reveals three distinct regimes, each with its own scaling behavior, indicating that no single universal friction law governs friction across this wide range of parameters. This finding is consistent with different droplet volumes and regardless of whether the surfaces are superhydrophobic, super-slippery  (black silicon), or superheated. Fig. \ref{fig:5} shows that there exist a regime where the normalized friction is independent of the capillary number (black-silicon: $Ca<5\times10^{-7}$ and Glaco-coated surface: $Ca<5\times10^{-6}$), indicating that this regime is dominated by contact-line friction: $F_T/2r\sigma \propto \cos\theta_r - \cos\theta_a $. When the capillary number is progressively increased, we find that the the non-dimensional friction scales as $Ca^{2/3}$ up to $Ca \approx 5\times 10^{-5}$ for both superhydrophobic and black-silicon surfaces where the droplet levitates over a thin film governed by Landau–Levich–Derjaguin law \cite{keiser2017drop,keiser2020universality} even for the oil droplet on a water layer cases of \citet{daniel2019hydration}.  
As the capillary number is further increased  $5\times 10^{-5}<Ca<7.5\times10^{-5}$ the slope of the experimental data deviates from $Ca^{2/3}$ (inset), indicating an emergence of an additional source of dissipation (term 1 in Eq.\ref{eq:1}) that becomes relevant at higher capillary number. 

A noticeable slope change and variation in collapse occurs at $Ca>9\times10^{-5}$ (Fig.\ref{fig:5}). The data in this region corresponds to a droplet moving on an inclined surface at terminal velocity, where aerodynamic drag is an additional source of friction as suggested by \citet{mouterde2019superhydrophobic}. The slope of the data trend reveals a scaling of $Ca^{4/3}$. However, we find that while the non-dimensional friction forces for two different droplet sizes (42 and 100 $\mu L$) and surfaces follow a similar slope, they do not fully collapse. At these velocities the proposed scaling law ($F_T/2r\sigma \approx Ca^{4/3}$) has an additional dependency on the droplet radius and surface characteristics, therefore, lacks universality to describe the friction force.

 \begin{figure}
\includegraphics{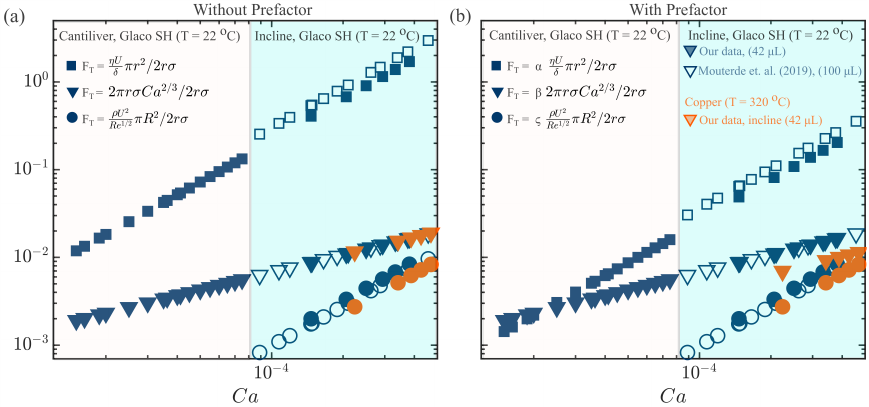}
\caption{\label{fig:6} Forces that make up the total force of friction (Eq.\ref{eq:1} \& \ref{eq:2}) as a function of capillary number, a) without a prefactor and b) with prefactors from Table~\ref{tab:table1}. Viscous dissipation in the bulk (squares), viscous dissipation from the droplet curvature (triangles), and aerodynamic drag (circles) as marked. Other data extracted from \cite{mouterde2019superhydrophobic}.  Cantilever experiments are dominated by viscous dissipation in the rim when $Ca<2\times10^{-4}$ and viscous dissipation in the bulk plays a larger role as $Ca$ which is easier to see in b). On the inclined plate, viscous dissipation in the bulk dominates the total friction forces, however, aerodynamic drag does have a higher slope, indicating that at higher $Ca$ aerodynamic drag may dominate. Viscous dissipation in the bulk is ignored for Leidenfrost cases because of the very thick vapor layer.}
\end{figure}

We plot the contributions of each term in Eq.\ref{eq:1} \& \ref{eq:2} as a function of the capillary number in Fig. \ref{fig:6} to analyze the contributions of each friction component, when levitated. Specifically, at higher droplet velocity (large capillary number), we find that there is significant contribution of the viscous dissipation in the droplet to the total friction, regardless of whether the droplet or the superhydrophobic surface is moving (Fig. \ref{fig:6}a). In Fig. \ref{fig:6}b the prefactors are added and it becomes more evident how the viscous dissipation term becomes more important as the capillary number is increased (increased velocity).  In the cantilever region the viscous dissipation term becomes more important $Ca^{2/3}>1\times 10^{-5}$, and in the inclined plate region the viscous dissipation and $Ca^{2/3}$ terms are much larger than the aerodynamic drag term. 

\begin{figure}
\includegraphics{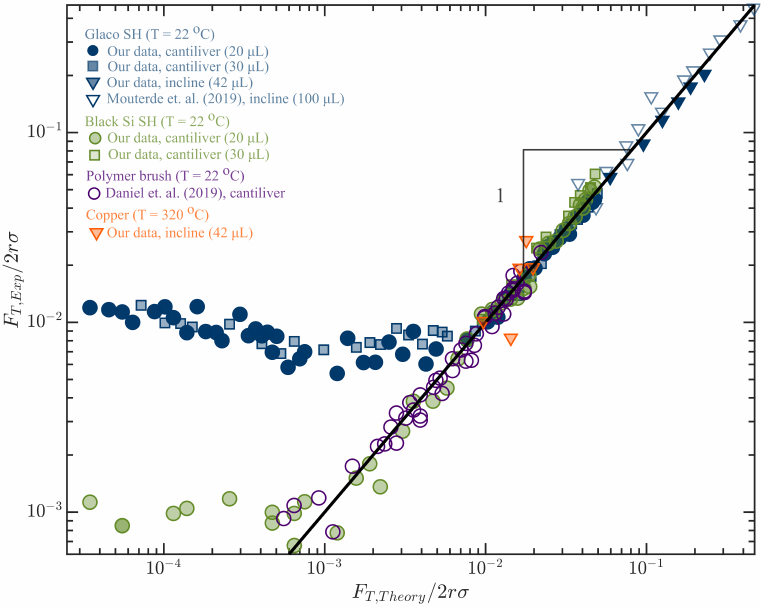}
\caption{\label{fig:7}Comparison of normalized experimental ($F_{T,Exp}$) and theoretical friction forces ($F_{T,Theory}$) based on  Eq.\ref{eq:1} \& \ref{eq:2}. Others data  extracted from \cite{mouterde2019superhydrophobic, daniel2019hydration}.}
\end{figure}

In response to this data, we propose a rather simple universal scaling law governed by Eqs.\ref{eq:1}\&\ref{eq:2}, which considers all of the forces for which the discrepancy seen in Fig. \ref{fig:5} vanishes, leading to a congruence between the experimental data and proposed scaling model, as visualized by Fig. \ref{fig:7}. The data from our superhydrophobic cases and \citet{mouterde2019superhydrophobic} collapse when all the forces are considered for the same pre-factors as shown in Table~\ref{tab:table1}. Note, the pre-factors used herin are different than those used by \cite{mouterde2019superhydrophobic} where they ignored the second term, and different than those used by \cite{daniel2019hydration}. 

\begin{table}
\caption{Prefactors used for estimating the forces at play in the droplet data presented. We apply the same prefactors for a given type of experiment to all cases included those extracted from other studies. However, we also present the prefactors  other studies used as a comparison. } 
\begin{tabular}{lccr}
\hline
\textrm{Force prefactor}&
\textrm{Cantilever}&\textrm{Incline superhydrophobic}&
\textrm{Incline Leidenfrost}\\
from Eq.\ref{eq:1}\&\ref{eq:2}&
\textrm{herein, others}&\textrm{herein, others}&\\

\colrule
$\alpha$ & 0.12, 0\cite{daniel2019hydration} & 0.12, 0.85\cite{mouterde2019superhydrophobic} & 0\\
$\beta$ & $\beta_1 = 6\pi$, 12\cite{daniel2019hydration} & $\beta_2=1$, 34\cite{mouterde2019superhydrophobic} & 0.6\\
$\zeta$ & 0 & 1 & 1\\
\hline
\label{tab:table1}
\end{tabular}
\end{table}

Furthermore, the Leidenfrost cases also scale with the forces outlined in Eq.\ref{eq:2}, however, the pre-factors are slightly different. Leidenfrost cases have much thicker vapour layers than the aerodynamic leidenfrost ones and thus the viscous dissipation term is ignored. 
The velocity of the Leidenfrost droplet near the pure rolling velocity in Fig.~\ref{fig:3}(a) also implies that the viscous dissipation of the droplet is near zero (i.e., $\alpha=0$). The viscous dissipation from the droplet curvature is overestimated in the Leidenfrost case as the thickness of the vapor layer is large, thus $\beta$ is decreased. Further, Landau-Levich-Derjaguin estimation can also be written as a function of the vapor thickness $(\eta_a U/h)2\pi r l \approx y(\eta_a U/\sigma)^{2/3}2\pi\sigma r$, where a thicker vapor layer will lower the frictional force. This can be accomplish by setting $\beta=0.6$. 



Although we are proposing a universal scaling for all droplet conditions, we are using all of the available forces that are discussed in the literature to do so. In particular, the model reveals that the dominant forces responsible for relative droplet motion between a substrate change from (i) contact angle hysteresis to (ii) viscous dissipation from droplet curvature to (iii) viscous dissipation in the bulk, and potentially (iv) aerodynamic drag at much higher velocities not reached in this study.



\section{\label{sec:interferometry}Thin film interferometry}

\begin{figure*}
\includegraphics{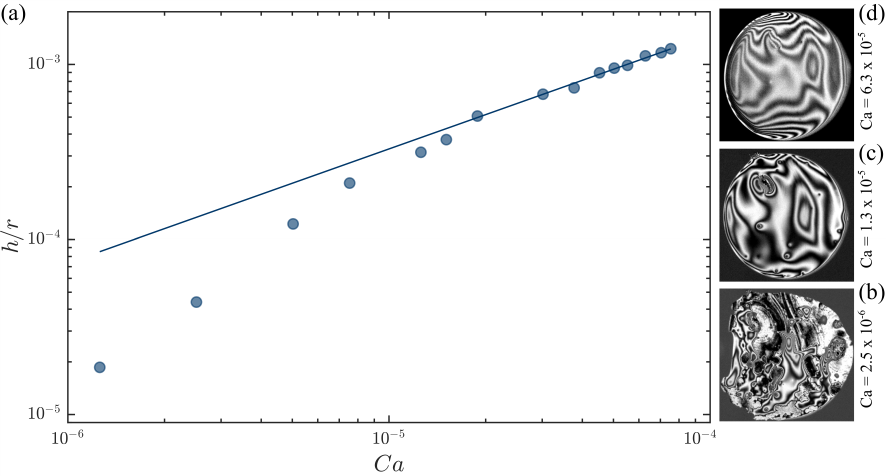}
\caption{\label{fig:8}(a) The air-film thickness in the central region of a 42 $\mu L$ water droplet, plotted as a function of the capillary number. The droplet moves at a controlled speed on a cantilever, and the air-film thickness is measured using white-light interferometry. The solid line represents the thickness predicted by the Landau-Levich-Derjaguin model, expressed as: $h \approx rCa^{2/3}$, where $Ca$ is the capillary number and $r$ is the contact radius. (b-d) Reflection interference contrast microscopy (RICM) is used to visualize change in the air film thickness when droplet moves from left to right on a superhydrophobic surface at progressively increased controlled velocity ($Ca$ as marked). RICM images show confirm the thin-air film thicknesses adhere to LLD law. (b) has a very small film that is in contact with the surface in many places, whereas (d) shows a larger air film that is thicker in the center with thinner portions top and bottom and more evenly distributed than the other two cases. }
\end{figure*}

Next, we show that the air-film thickness beneath the droplet increases nonlinearly with an increase in the velocity of a droplet moving on a superhydrophobic surface. We employ white-light interferometry to measure the air-film thickness around the flattest region of the droplet which falls along the mid-section of the droplet (shown in Fig. \ref{fig:8}a). We also perform time-resolved RICM imaging from below to establish correlations with friction forces stemming either form the contact line pinning or viscous dissipation (Fig. \ref{fig:8}b-d). We find that for a specific range of capillary numbers ($10^{-5} < Ca < 10^{-4}$), where friction scales as ($Ca^{2/3}$), the air-film thickness beneath the droplet varies in accordance with $h \approx Ca^{2/3}$, as seen in Fig. \ref{fig:8}(a). This observation suggests that the air-film thickness adheres to the Landau-Levich-Derjaguin (LLD) law \cite{landau1988dragging}, which correlates with the absence of contact line pinning \cite{daniel2017oleoplaning}, typically observed when a moving droplet entrains air from its surroundings. The entrained air generates a lift force, causing the droplet to levitate above the surface \cite{daniel2017oleoplaning} (aka aerodynamic Leidenfrost). The LLD theory breaks down when $Ca$ drop below $10^{-5}$. This is evidenced in Fig. \ref{fig:8}(a) where the thickness of air-film does not seem to scale with $Ca^{2/3}$.

\par We perform RICM imaging for a broad range of capillary numbers to find the origin of the friction for a droplet moving over tilted superhydrophobic surface. Fig. \ref{fig:8}(d-f) show RICM images for three different capillary numbers, each representing distinct dynamics. We find that a droplet moving on superhydrophobic surface at a low velocity, $U=0.01$ ms$^{-1}$ ($Ca = 2.5\times10^{-6}$) gives rise to heterogeneous wetting, as seen in Fig \ref{fig:8}(b). This suggests that the friction force affecting the droplet motion primarily stems from contact line pinning \cite{schellenberger2015direct,gao2018drops}. This is consistent with the observation made in Fig. \ref{fig:4} at $Ca=2.5\times10^{-6}$ where friction forces are independent of velocity. Increasing the velocity beyond $Ca>1.3\times10^{-5}$ increases the air-film thickness as shown in the RICM image (c). However, formation of capillary bridges at the receding contact line and presence of discrete pinning/de-pinning points suggests that friction originates from both contact line friction and viscous dissipation around the rim \cite{daniel2018origins}. Further increasing the droplet velocity reveals a  smooth and continuous receding and advancing front with no presence of capillary bridges as seen in Fig. \ref{fig:8}d. Furthermore, the absence of visible discrete pinning points provides a corroborative evidence that the droplet has levitated.  

\section{\label{sec:Conclusion}Conclusion}
We have investigated the frictional forces acting on droplets moving across superhydrophobic and superheated surfaces. Experiments consisted of surfaces placed on an inclined substrate ($\theta = 0$\textdegree to 6\textdegree) or a droplet in contact with a cantilever force sensor on a moving substrate. High-speed interferometric visualization and RICM imaging were used to show the thickness of the air-layer ($h$) beneath the droplet.

Our findings reveal that the frictional forces vary significantly based on the surface type and the dynamic behavior of the droplet. On superhydrophobic surfaces, the friction force is nearly constant and dominated by contact-angle hysteresis at low capillary numbers. Increasing the capillary number results in an increase in the viscous dissipation around the droplet rim curvature following a $Ca^{2/3}$ scaling law. As the capillary number is increased further the $Ca^{2/3}$ scaling begins to break down as viscous dissipation in the bulk plays a larger role even in the cantilever experiments ($Ca>10^{-5}$). Inclined surfaces are used to achieve $Ca>10^{-4}$ where viscous dissipation in the bulk begins to dominate with forces scaling closer to $Ca^{4/3}$.  In contrast, droplets on superheated surfaces (Leidenfrost effect) at similar capillary numbers achieve much lower friction values.

In response to these observations we propose a universal scaling law (Eqs. \ref{eq:1} \& \ref{eq:2}) that incorporates all relevant forces—contact angle hysteresis, viscous dissipation from droplet curvature near the substrate, viscous dissipation in the bulk, and aerodynamic drag—providing a more comprehensive understanding of droplet dynamics on non-wetting surfaces. This model accurately predicts the transitions between different frictional regimes, contributing to a better grasp of the mechanisms governing droplet motion.

The insights gained from this research hold significant implications for the design of non-wetting surfaces in various applications, from fluid transport systems to microfluidic devices. By elucidating the underlying frictional forces, our study paves the way for optimizing surface designs to minimize friction and enhance efficiency in relevant technological processes.

\begin{acknowledgments}
We wish to thank KAUST for funding, Dilip Maity for insightful discussions and Addison Litton for inspiring this follow on study.
\end{acknowledgments}

\appendix


\bibliography{apssamp}

\end{document}